\documentstyle[twocolumn,psfig,aps]{revtex}

\begin{document}
\twocolumn[\hsize\textwidth\columnwidth\hsize\csname
@twocolumnfalse\endcsname
\title{Coexistence of charge density wave and spin-Peierls orders in 
quarter-filled quasi-one dimensional correlated electron systems}
\author{J.~Riera$^{a,b}$ and D.~Poilblanc$^a$}
\address{
$^a$Laboratoire de Physique Quantique \& UMR--CNRS 5626,
Universit\'e Paul Sabatier, F-31062 Toulouse, France\\
$^b$Instituto de F\'{\i}sica Rosario, Consejo Nacional de
Investigaciones 
Cient\'{\i}ficas y T\'ecnicas, y Departamento de F\'{\i}sica,
Universidad Nacional de Rosario, Avenida Pellegrini 250, 2000-Rosario,
Argentina
}
\date{\today}
\maketitle

\begin{abstract}
Charge and spin-Peierls instabilities in quarter-filled ($n=1/2$)
compounds consisting of coupled ladders and/or zig-zag chains
are investigated. Hubbard and t-J models including local Holstein 
and/or Peierls couplings to the lattice are studied by
numerical techniques. Next nearest neighbor hopping and magnetic 
exchange, and short-range Coulomb interactions are also considered.
We show that, generically, these systems undergo instabilities
towards the formation of Charge Density Waves, Bond Order Waves 
and (generalized) spin-Peierls modulated structures.
Moderate electron-electron and electron-lattice couplings can lead to
a coexistence of these three types of orders.  
In the ladder, a zig-zag pattern is stabilized by the Holstein
coupling and the nearest-neighbor Coulomb repulsion. In the case of an
isolated chain, bond-centered and site-centered $2k_F$ and $4k_F$
modulations are induced by the local Holstein coupling. 
In addition, we show that, in contrast to the ladders,
a small charge ordering in the chains, strongly enhances 
the spin-Peierls instability. Our results are applied to the
NaV$_2$O$_5$ compound (trellis lattice) and various phases with
coexisting charge disproportionation and spin-Peierls
order are proposed and discussed in the context of recent experiments. 
The role of the long-range Coulomb potential is also outlined.

\end{abstract}

\vskip2pc]


\section{Introduction}

Quasi-one dimensional correlated electrons systems at 
the commensurate filling of $n=1/2$ (quarter-filled band) 
show fascinating physical properties. 
A widely studied class of 
such materials are the so-called organics
charge transfer salts like the Bechgaard salts~\cite{review}.
These systems which consist of stacks of organics molecules forming
weakly coupled one dimensional chains 
exhibit, at low temperature, a large variety of exotic phases 
such as superconducting, spin density wave, charge density wave (CDW)
and spin-Peierls (SP) phases. 

The vanadium inorganic compound NaV$_2$O$_5$ is also believed to be 
a nearly perfect realization of a quarter-filled low dimensional
system. Therefore, the nature of the SP phase~\cite{discovery,Fertey}
below $T_{SP}\simeq 35K$ 
is expected to be quite different from the one occurring
in the more conventional antiferromagnetic Heisenberg chain 
CuGeO$_3$ (Ref.~\cite{CuGeO3}).

\begin{figure}
\begin{center}
\psfig{figure=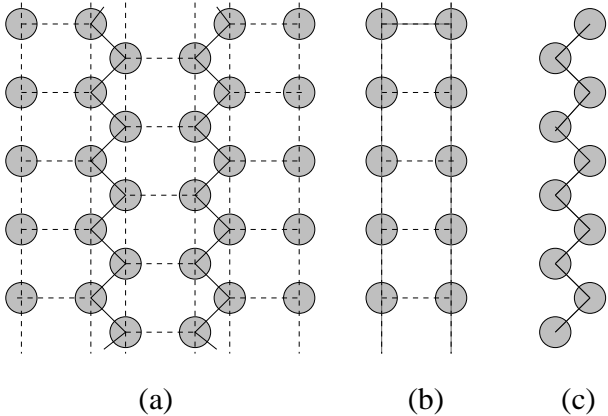,width=8truecm,angle=0}
\end{center}
\caption{
(a) Structure of the vanadium planes of NaV$_2$O$_5$ 
showing ladder (b) and chain (c)
patterns. Each vanadium carries, on average, a charge of $+0.5$. 
}
\label{NaVO}
\end{figure}

The NaV$_2$O$_5$ system is built from weakly coupled planes whose
structure is shown in Fig.~\ref{NaVO}. It can be depicted as
an array of parallel ladders (Fig.~\ref{NaVO}(b)) coupled in
a trellis lattice. Note that a small buckling 
of the V plane can be neglected in first approximation.
Oxygen atoms (not shown) are located at the center of the vertical 
and horizontal bonds
of Fig.~\ref{NaVO}(a) and lead to effective hopping matrix elements
and antiferromagnetic (AF) super-exchange interactions. 
LDA band structure calculations~\cite{LDA} and estimations 
based on empirical rules~\cite{empirical} 
lead to similar values of the hopping amplitudes along and
perpendicular to the ladders, $t_\parallel\simeq 0.15{\rm eV}$ 
and $t_\perp\simeq 0.35{\rm eV}$ respectively. However, 
some controversy remains regarding the magnitude of the
diagonal hopping (see Fig.~\ref{NaVO}(c)) $t_{xy}$ with values ranging 
from $0.012\rm{eV}$ (Ref.~\cite{LDA}) to $0.3{\rm eV}$ 
(Ref.~\cite{empirical}).

Although the average valence of the vanadium in NaV$_2$O$_5$ is 4.5
(half an electron per vanadium d-orbital on average),
the exact nature of the charge ordering is still under active debate. 
Early X-rays diffraction experiments~\cite{Galy} were pointing in
favor of a non-centrosymmetric structure implying two inequivalent
vanadium sites. It was further suggested~\cite{Galy} that magnetic 
spin-1/2 $V^{4+}$ were forming one-dimensional (1D) chains separated by
non-magnetic $V^{5+}$ chains. Based on this assumption, various
theoretical analyses of this material were attempted using 
dimerized Heisenberg chains~\cite{dimerized_chain} or 
spin-phonon models~\cite{spin_phonon}.

Recently, a new structure refinement at room
temperature~\cite{Damascelli}
suggested that, in contrast to earlier reports~\cite{Galy},
$\alpha'$-NaV$_2$O$_5$ would have a centrosymmetric space group
implying only one kind of vanadium site. 
Besides, below the transition temperature $T_{SP}$,
joint neutron and X-ray diffraction experiments~\cite{LLB_Synchrotron} 
reveal new superlattice reflections which can be ascribed to a
lattice modulation associated to displacements of predominantly V atoms.

The insulating character of
NaV$_2$O$_5$ could, in fact, be simply understood in the framework of 
the quarter-filled Hubbard or
t--J ladders~\cite{empirical,photoemission,Nishimoto} without 
invoking any charge order mechanism (as it is expected at room
temperature). A simple analytic
picture, valid when the hopping along the legs is 
small~\cite{empirical,photoemission}, gives a finite charge gap
of the order of (for a very large Hubbard repulsion U)
the splitting $2t_\perp$ between the bonding and antibonding states
of the rungs.
The existence of a metal-insulator transition in quarter-filled
t--J ladder has been confirmed by numerical calculations
of the single particle density of state~\cite{photoemission} and
the charge gap~\cite{Nishimoto} for realistic parameters.
In the insulating state, the ground state (GS) configuration 
corresponds to the single occupancy of the bonding states on 
each rung and it has been argued
that, in this case, the low-energy processes can be described by an
effective AF Heisenberg chain~\cite{empirical,Nishimoto},
hence giving some relevance to earlier descriptions of this
material~\cite{dimerized_chain,spin_phonon}.
Furthermore, angle-resolved photoemission data 
at room temperature~\cite{Kobayashi} are consistent both with
a description in term of an effective half-filled 
t-J chain~\cite{dimerized_chain} or in terms of a quarter-filled 
t-J ladder~\cite{photoemission}.

Recently, motivated by the experimental studies of NaV$_2$O$_5$
at low temperature, charge ordering has been investigated 
theoretically~\cite{Fukuyama,Fulde,Nishimoto,Khomskii}. It has been 
argued that repulsion between electrons on
neighboring ladders can lead to charge disproportionations
where the electrons are localized on a single leg
of each ladder~\cite{Nishimoto}
(as in the original structure proposed in Ref.~\cite{Galy})
or form a zig-zag pattern within the 
ladder~\cite{Fukuyama,Fulde,Khomskii}. Although the Madelung energy,
which includes also a long-range (LR) Coulomb interaction, slightly
favors the chain charge ordering, it has been suggested~\cite{Khomskii}
that, in contrast, the underlying coupling with the lattice could 
stabilize the zig-zag CDW.
Experimental features in the optical conductivity at intermediate 
energies (0.6--2.5eV)~\cite{charged_magnon} could be reproduced in
a calculation of the t-J model on the trellis lattice~\cite{Nishimoto2}
only by assuming an {\it ad-hoc} charge disproportionated GS, although 
low-energy charged magnons~\cite{charged_magnon} could not be found.

In this paper, our aim is to investigate further, by numerical
exact diagonalization (ED) techniques, the interplay between 
the electron-electron Coulomb repulsion and the electron-phonon
(or spin-phonon) couplings. Indeed, charge disproportionation
could come from a Holstein coupling, possibly related to the apical
oxygen, as well as from a LR Coulomb repulsion. 
Although the emphasis is put here on the understanding of the physical
properties of NaV$_2$O$_5$, we believe our results can also be
relevant for organic salts.
The likely smallness of the interladder as well as many of the 
experimental results point to the possibility that the charge or SP
instabilities are driven primarily by the ladder
physics.~\cite{LDA,Khomskii}
Thus, the first step of our analysis consists of the study of these 
structures by considering isolated quarter-filled t-J 
ladders including lattice-local charge couplings 
(Holstein) or modulations of the bond parameters (Peierls and SP
couplings).
We shall then try to determine whether the states found for the ladders
are also consistent with the physics of the chains. 
Eventually, we shall consider
the trellis lattice (Fig.~\ref{NaVO}(a)) of NaV$_2$O$_5$
where the most likely states will be determined based on the
previous results obtained for the individual ladder and zig-zag chain 
sublattices.

\section{Isolated ladder: charge instability}

We shall first consider the quarter-filled anisotropic
t-J ladder (Fig.~\ref{NaVO}(b)) in the presence of an Holstein-coupling,
\begin{eqnarray}
H&=&H_{tJ}+H_V+H_H \label{Ham1} \\
H_{tJ}&=&J_\parallel \sum_{i,\alpha} ({\bf S}_{i,\alpha}\cdot
{\bf S}_{i+1,\alpha}-\frac{1}{4}n_{i,\alpha}n_{i+1,\alpha})
\nonumber \\
&+&J_\perp \sum_{i} ({\bf S}_{i,1}\cdot {\bf S}_{i,2}
-\frac{1}{4}n_{i,1}n_{i,2})\nonumber \\
&+&t_\parallel \sum_{i,\alpha,\sigma}
({\tilde c}_{i,\alpha;\sigma}^\dagger{\tilde c}_{i+1,\alpha;\sigma}
+ h.c.)
\nonumber \\
&+&t_\perp \sum_{i,\sigma} ({\tilde c}_{i,1;\sigma}^\dagger
{\tilde c}_{i,2;\sigma}+h.c.) \nonumber \\
H_V&=& V\sum_{i} (n_{i,1}n_{i+1,1}+n_{i,2}n_{i+1,2}+n_{i,1}n_{i,2})
\nonumber \\
H_H&=& \sum_{i,\alpha} n_{i,\alpha}\delta_{i,\alpha} 
+ \frac{1}{2}K\sum_{i,\alpha}\delta_{i,\alpha}^2
\nonumber \, ,
\end{eqnarray}
where 
${\tilde c}_{i,\alpha;\sigma}^\dagger
= c_{i,\alpha;-\sigma}(1-n_{i,\alpha;\sigma})$ 
are {\it hole} Guzwiller projected creation operators (the large
on-site Coulomb interaction prevents doubly-occupancy) and the
index $\alpha$ stands for a chain index ($=1,2$). We consider
{\it a priori} different fermion hopping amplitudes ($t_\parallel$,
$t_\perp$) or magnetic exchange interactions ($J_\parallel$,
$J_\perp$) along the legs
and along the rungs. Hamiltonian (\ref{Ham1}) can be viewed as the
strong coupling limit of a Hubbard ladder so that one can assume
a relation of the form $J_\parallel/J_\perp=(t_\parallel/t_\perp)^2$ 
between the parameters. A nearest neighbor (NN) Coulomb repulsion $V$
has been included. Note that the electron-phonon coupling has been
absorbed in the definition of the lattice displacement $\delta_i$.
The magnitude of the coupling to
the lattice is then given by a single parameter namely
the inverse of the lattice stiffness $1/K$.
The phonons have been given an infinite mass since the 
charge and spin dynamics are assumed here to involve smaller time
scales than lattice fluctuations (adiabatic approximation).
Note that the on-site displacement $\delta_i$ corresponds in fact
to an effective parameter which might combine several
effects. Physically, it could correspond to
the displacement of the apical oxygen and/or to 
the displacements of the neighboring in-plane oxygen atoms toward or 
away from the V site. 

Before proceeding with the study of Hamiltonian
(\ref{Ham1}), it is instructive to recall the properties of the
t-J ladder in the absence of electron-phonon coupling and
NN repulsion. The electronic properties of this model at quarter
filling have been investigated
previously~\cite{empirical,photoemission,Nishimoto} and the existence
of a metal-insulator transition has been
shown~\cite{photoemission,Nishimoto}.
It is believed that the system becomes insulating for
(approximately) $t_\perp> 2t_\parallel$. Physically,
this corresponds to the situation where the bonding and antibonding
bands are completely separated, the lower band becoming effectively
half-filled so that an arbitrary small repulsion opens a charge
gap. This physical situation might be relevant for the insulating phase
of the NaV$_2$O$_5$ material.

Although the lattice is considered here in the adiabatic approximation,
no supercell order is assumed {\it a priori} and the lowest energy
equilibrium lattice configuration is obtained through a self-consistent
procedure (as in the rest of the paper).
Indeed, the total energy functional $E(\{\delta_{i,\alpha}\})$
can be minimized with respect to the sets of distortions
$\{\delta_{i,\alpha}\}$ by solving the non-linear set of local
coupled equations,
\begin{equation}
K\delta_{i,\alpha}+\big<n_{i,\alpha}\big>=0\, ,
\label{non_linear}
\end{equation}
where $\big<...\big>$ is the GS mean value obtained by ED (using
the Lanczos algorithm) of Hamiltonian (\ref{Ham1}). Since the second
term depends implicitly on the distortion pattern
$\{\delta_{i,\alpha}\}$, Eqs.~(\ref{non_linear}) can be solved by a
regular iterative procedure.\cite{Dobry_Riera}

The phase diagram of the Holstein-t-J ladder model is shown 
in Fig.~\ref{ladder_Holstein}(a) for a realistic set of parameters.
These results have been obtained by studying $2\times 6$ and 
$2\times 8$ clusters.
In the absence of NN repulsion $V$, a rapid transition occurs
from a uniform phase (U) at small coupling (or equivalently large
lattice rigidity) to a localized phase at large coupling.
This strong coupling phase is characterized by a
charge ordering with two types of (almost) completely empty or
completely occupied sites arranged in some disordered patterns. 
More interestingly, a CDW phase with a ``zig-zag'' arrangement of the
excess charge (which exists also for $V=0$ only in a very narrow region 
around $1/K\sim 2.5$) is stabilized by the NN repulsion $V$. 
In contrast to the localized phase, the charge disproportionation in
this zig-zag CDW state is not complete. Notice that there is a finite
critical value of $1/K$ associated to the 
stability of the CDW phase. This feature might be due to the fact that
the uniform ladder for $1/K=0$ is, for an anisotropy ratio of
$t_\parallel/t_\perp=0.4$, already in an insulating state with a
charge gap (see Ref.~\cite{photoemission}).

\section{Isolated ladder: coexisting charge and SP orders}

The possibility of a coexisting SP order in the previous CDW state can
be studied by considering additional Peierls and SP couplings realized
by making the following substitutions in Hamiltonian (\ref{Ham1});
\begin{eqnarray}
t_{\parallel} \,\rightarrow\, t_{\parallel}
(1+\delta^B_{i,\alpha})\, ,
\label{Peierls_ladder1} \\
J_{\parallel} \,\rightarrow\, J_{\parallel}(1+g\delta^B_{i,\alpha})\, ,
\label{Peierls_ladder2}
\end{eqnarray}
and by adding a new elastic term 
$\frac{1}{2}K_B\sum_{i,\alpha}(\delta_{i,\alpha}^B)^2$.

Note that the spin-Peierls order is, for a quarter-filled band,
intrinsically linked to a bond order wave (BOW) characterized by a
modulation of the hopping amplitudes~\cite{note_BOW} as given by
Eq.~(\ref{Peierls_ladder1}).
In other words, the expectation values
$\big<c_{i,\alpha;\sigma}^\dagger c_{i+1,\alpha;\sigma}\big>$ and
$\big<{\bf S}_{i,\alpha}\cdot {\bf S}_{i+1,\alpha}\big>$  should
exhibit similar modulations along the legs. For simplicity, 
we shall assume here that the magnetic exchange interation
$J_\parallel$ involves two virtual hops $t_\parallel$ so that one can
take, in first approximation, $g\simeq 1$. However, the qualitative 
results of this study do not depend on the choice of $g$. 
It should be noticed in addition that we shall consider modulations
only along the legs of the ladders
since our analysis showed that modulations of the rungs were not
favored.

\begin{figure}
\begin{center}
\psfig{figure=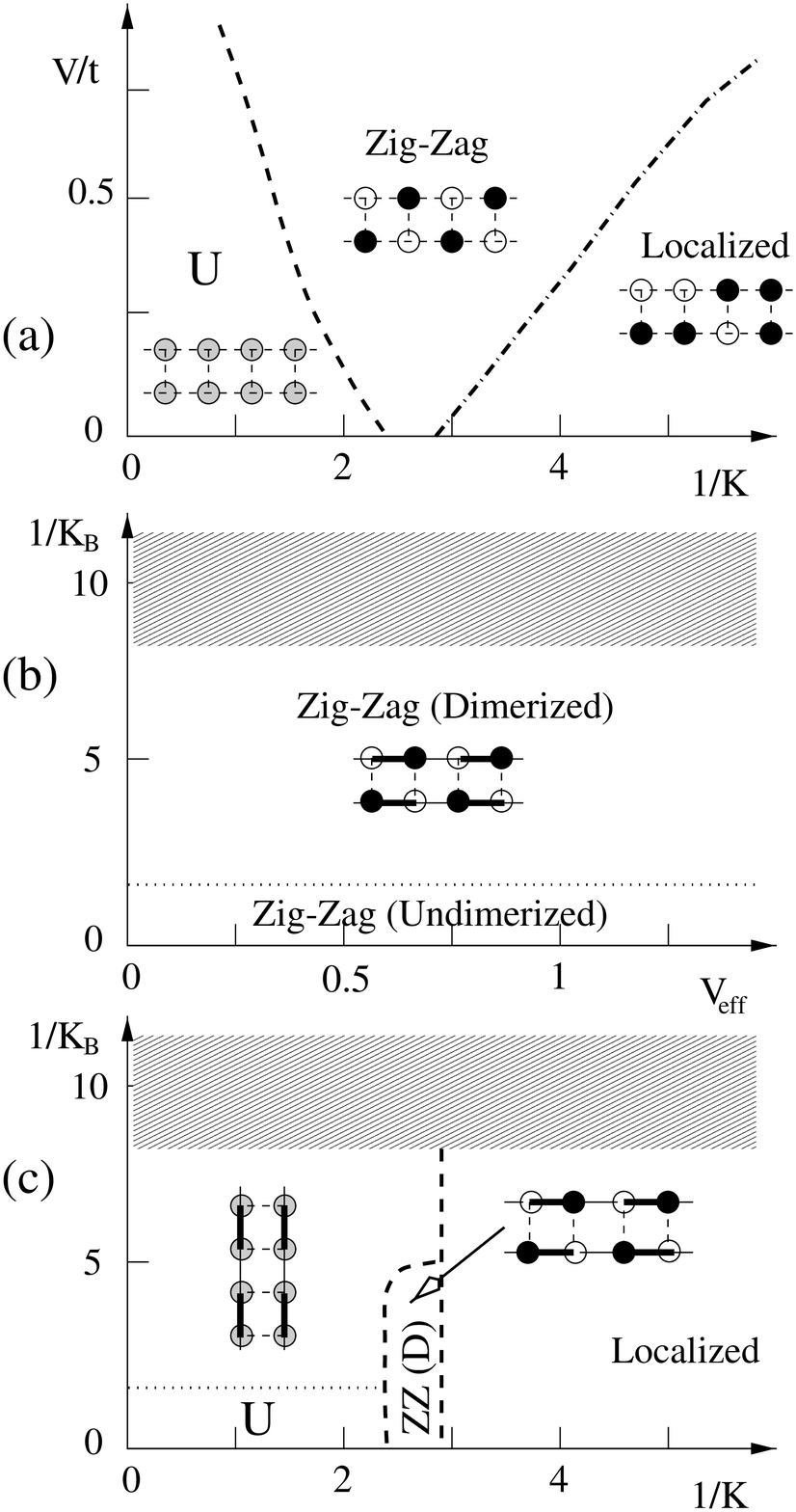,width=8truecm,angle=0}
\end{center}
\caption{
Typical phase diagrams of $\frac{1}{4}$-filled anisotropic t-J ladders 
(a) as a function of the NN Coulomb repulsion $V$ 
and the on-site Holstein coupling of strength $1/K$; 
(b) as a function of the Peierls lattice coupling and the effective 
potential $V_{\rm eff}$;
(c) as a function of Holstein and Peierls
lattice couplings.  
These results have been obtained by ED
of small clusters with anisotropy ratios $t_\parallel/t_\perp=0.4$
and $J_\parallel/J_\perp=0.16$.
Shaded regions are unphysical (see text).
}
\label{ladder_Holstein}
\end{figure}

\begin{figure}
\begin{center}
\psfig{figure=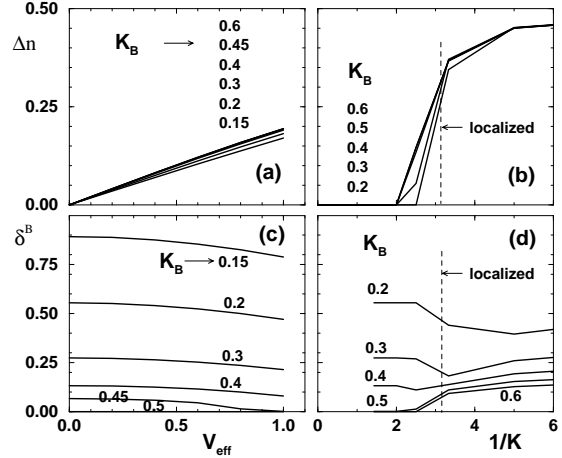,width=8truecm,angle=-90}
\end{center}
\caption{
CDW order parameter $\Delta n$ (a,b)
and bond modulation $|\delta^B_i|$ (c,d) of a $\frac{1}{4}$-filled 
t--J ladder for various lattice bond stiffness $K_B$ (as
indicated on the plots).
(a,c) as a function of the effective potential $V_{\rm eff}$;
(b,d) as a function of the on-site Holstein coupling strength $1/K$. 
In the ``localized'' phase (on the right of the dashed line),
estimates based on an average over the system are shown.
These results have been obtained by ED
of small clusters with anisotropy ratios $t_\parallel/t_\perp=0.4$
and $J_\parallel/J_\perp=0.16$.
}
\label{plot_ladder}
\end{figure}

Since our purpose is to investigate the role of the additional Peierls
coupling on the zig-zag CDW state, we shall first consider an
{\it ad-hoc} effective potential,
\begin{equation}
H_{\rm eff} = V_{\rm eff} \sum_{i,\alpha}
(-1)^{(x_{i,\alpha}+y_{i,\alpha})} n_{i,\alpha}\, ,
\end{equation}
replacing the terms $H_H+H_V$ in Hamiltonian (\ref{Ham1})
in order to stabilize the zig-zag pattern.
Qualitatively, this term mimics the combined effects of
the Holstein lattice coupling and the NN Coulomb repulsion.
The previous method based on Eq.~(\ref{non_linear}) can be
extended here to deal with the lattice-bond interaction. The
lowest energy configuration is found by minimizing the GS energy
with respect to the full set of parameters $\{ \delta^B_{i,\alpha}\}$.
This is achieved by solving exactly, using
the same iterative procedure as before, the set of
non-linear equations, 
\begin{eqnarray}
K_B\delta^B_{i,\alpha} &+& gJ_\parallel \big<{\bf S}_{i,\alpha}\cdot
{\bf S}_{i+1,\alpha}-\frac{1}{4}n_{i,\alpha}n_{i+1,\alpha}\big>
\nonumber \\
&+&t_\parallel
\big< {\tilde c}_{i,\alpha;\sigma}^\dagger{\tilde c}_{i+1,\alpha;\sigma}
+ h.c.\big> =0\, .
\label{non_linear2}
\end{eqnarray}
on finite clusters.
As shown in the phase diagram of Fig.~\ref{ladder_Holstein}(b),
a finite value of the coupling strength $1/K_B\sim 2$ is needed to
produce a dimerized pattern overimposed to the CDW state.  
Above this critical value, a finite modulation of the form
$\delta^B_{i,\alpha}
=\delta^B (-1)^{x_{i,\alpha}}$
(the ladder is oriented along the x-axis) appears simultaneously with
the CDW order parameter for any $V_{\rm eff}$.
As can be seen in Fig.~\ref{plot_ladder}(a),
the charge disproportionation,
$\Delta n = |\big< n_i\big> -\frac{1}{2}|$,

\noindent
increases almost
linearly with the magnitude of the effective potential and depends
weakly on the magnetoelastic coupling as expected. On the other hand,
the amplitude $\delta_B$ of the dimerization
(Fig.~\ref{plot_ladder}(c)) increases strongly with $1/K_B$
when $1/K_B>2$.
However, for very strong couplings, our method gives an
unphysical solution $\delta_B>1$ (implying ferromagnetic bonds)
shown by a shaded region in Fig.~\ref{ladder_Holstein}(b)
(as in the rest
of the paper). This signals that our simple initial model
breaks down for such a strong coupling with the lattice and that
other terms should be included (quartic terms, etc...). 
Interestingly enough, the presence of a zig-zag charge
pattern is not at all required to observe a dimerization in the ladder.
On the contrary, as it is clear in Fig.~\ref{plot_ladder}(c),
the dimerization occurs for $V_{\rm eff}=0$ (i.e. for a uniform
distribution of the charge) and is even weakly {\it suppressed} by
$V_{\rm eff}$.

We now turn to the investigation of the full Hamiltonian
including both Holstein and Peierls-like lattice couplings, the
amplitudes of each of them being characterized by the two independent
coupling constants $1/K$ and $1/K_B$ respectively. 
The GS configuration for the site and bond deformations can be
obtained exactly on small clusters by solving simultaneously, by
the previous recursive method, the two sets of non-linear
equations (\ref{non_linear}) and (\ref{non_linear2}).
As before, at each step of the recursive procedure, the
quantum mechanical Hamiltonian is diagonalized by using a Lanczos
algorithm. Our results, for the same set of parameters $t_\mu$,
$J_\mu$ as above, are summarized in the phase diagram of
Fig.~\ref{ladder_Holstein}(c) with the corresponding behaviors of the
local CDW order
parameter and bond deformation shown in Figs.~\ref{plot_ladder}(b,d).
In contrast to the previous model, no zig-zag CDW ordering appears at
small Holstein coupling $1/K$. A uniform dimerized (D) phase
is stable for large enough $1/K_B$. It is important to notice
that the dimerized zig-zag (D-ZZ) phase is
found only in a narrow region located around $1/K\sim 2.5$ while
a larger Holstein coupling stabilizes a localized phase (with
no well-defined periodic pattern). In addition, these data
suggest that the Peierls coupling rather tends to suppress the
charge order and hence to destabilize the D-ZZ phase with respect
to the D phase. However, the region of stability of the D-ZZ phase
(of particular interest in the case of the NaV$_2$O$_5$ material) is
expected to be greatly enhanced by a NN Coulomb repulsion
(see Fig.~\ref{ladder_Holstein}(a))
as we have checked numerically.

\section{Isolated chain: charge instability}

\begin{figure}
\begin{center}
\psfig{figure=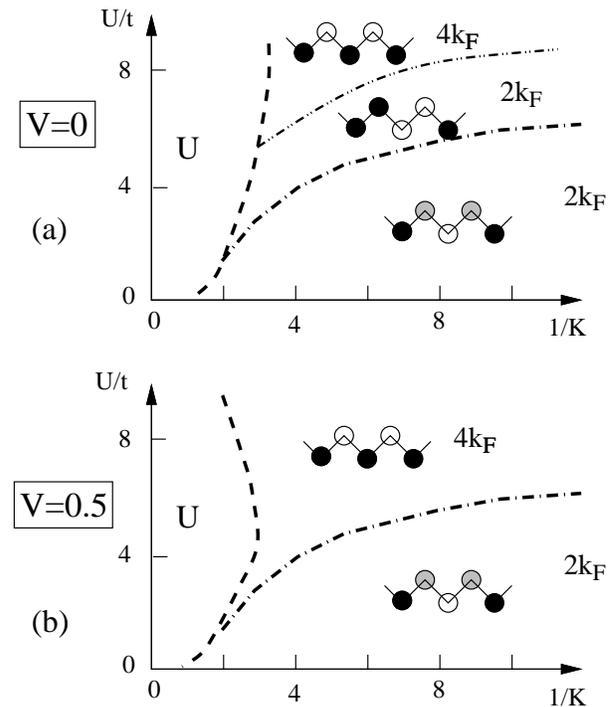,width=8truecm,angle=0}
\end{center}
\caption{
Typical phase diagrams of a $\frac{1}{4}$-filled 
Hubbard chain in the presence of an on-site Holstein coupling
without (a) or with a nearest neighbor Coulomb repulsion $V$ (b)
obtained from ED of small periodic chains. 
}
\label{Hubbard_Holstein}
\end{figure}

As mentioned above, in NaV$_2$O$_5$, the ladders are coupled by
diagonal bonds forming zig-zag chains. Besides, as mentioned in the
Introduction, many quasi-1D molecular
compounds contains weakly coupled chains. Therefore, we shall now
consider the role of the Holstein and Peierls lattice couplings on
one-dimensional chains. For completeness, we start with a
1D Hubbard model at quarter-filling ($n=1/2$)
coupled with an on-site (classical) phonon field,
\begin{eqnarray}
H=&t& \sum_{i,\sigma} (c_{i;\sigma}^\dagger
c_{i+1;\sigma}+h.c.)+U\sum_{i} n_{i;\uparrow}n_{i;\downarrow}
\nonumber \\
&+& V\sum_{i} n_{i}n_{i+1}
+ \sum_{i} n_{i}\delta_{i} + \frac{1}{2}K\sum_{i}\delta_{i}^2 \, .
\label{Ham2}
\end{eqnarray}
At quarter-filling, the Fermi wave vector is given by
$q_{2k_F}=\frac{\pi}{2}$ so that, at small $U$, one expects an
instability towards a 
$2k_F$-CDW state of wavevector $\lambda_{2k_F}=4a$ ($a$ is the
lattice spacing) mediated by the electron-phonon coupling. 
In contrast, for large U, the system becomes more similar to a
gas of interacting spinless fermions and the instability is
likely to occur at wavevector $2k_F^{SF}=4k_F$.
More generally,
we can parametrize the charge density as,\cite{note_BOW}
\begin{eqnarray}
\big< n_i\big>=\frac{1}{2}&+&A_{2k_F} \cos{(2\pi\frac{r_i}{4a}+
\Phi_{2k_F})}
\nonumber \\
&+&A_{4k_F} \cos{(2\pi\frac{r_i}{2a}+\Phi_{4k_F})}\, .
\label{charge_density}
\end{eqnarray}
A schematic phase diagram is shown in
Fig.~\ref{Hubbard_Holstein}(a) for $V=0$.
The uniform U metallic phase ($A_{2k_F}=A_{4k_F}=0$) is restricted to
a region at small electron-phonon coupling.
Above a critical line $1/K$ vs $U$, three different insulating 
CDW phases can be distinguished; (i) at small $U$, a $2k_F$-CDW phase
($A_{4k_F}\simeq 0$) centered on the sites, i.e. with $\Phi_{2k_F}=0$;
(ii) at intermediate $U$ (in the range 4--8), a {\it
bond}--centered $2k_F$-CDW phase i.e. with $\Phi_{2k_F}=\frac{\pi}{4}$;
(iii) at large U, a $4k_F$-CDW ($A_{2k_F}\simeq 0$, $\Phi_{4k_F}=0$).
As seen in Fig.~\ref{Hubbard_Holstein}(b), a small NN repulsion
suppresses completely the intermediate phase and enlarges the
region of stability of the $4k_F$-CDW phase~\cite{note_4kF}.

We now restrict ourselves to the strong electron correlation limit
(t-J model) to discuss further the role of the NN repulsion $V$.
Besides, since in NaV$_2$O$_5$ the zig-zag chains form part of the
trellis lattice we include 
NNN hopping matrix elements $t'$ and exchange integrals $J'$
which correspond to the interactions along the legs of
the ladders.  Hence, the Hamiltonian is,
\begin{eqnarray}
H&=&H_{tJ}+H_V+H_H     \label{tJ_chain}   \\
H_{tJ}&=&J \sum_{i} ({\bf S}_{i}\cdot {\bf S}_{i+1}-\frac{1}{4}
n_{i}n_{i+1}) +t \sum_{i,\sigma}
({\tilde c}_{i;\sigma}^\dagger{\tilde c}_{i+1;\sigma}
+ h.c.) \nonumber \\
&+&J' \sum_{i} ({\bf S}_{i}\cdot {\bf S}_{i+2}-\frac{1}{4}n_{i}n_{i+2})
+t' \sum_{i,\sigma} ({\tilde c}_{i;\sigma}^\dagger
{\tilde c}_{i+2;\sigma}+h.c.) \nonumber \\
H_V&=& V\sum_{i} n_{i}n_{i+1} \nonumber \\
H_H&=& \sum_{i} n_{i}\delta_{i} 
+ \frac{1}{2}K\sum_{i}\delta_{i}^2 \,. \nonumber 
\end{eqnarray}

\begin{figure}
\begin{center}
\psfig{figure=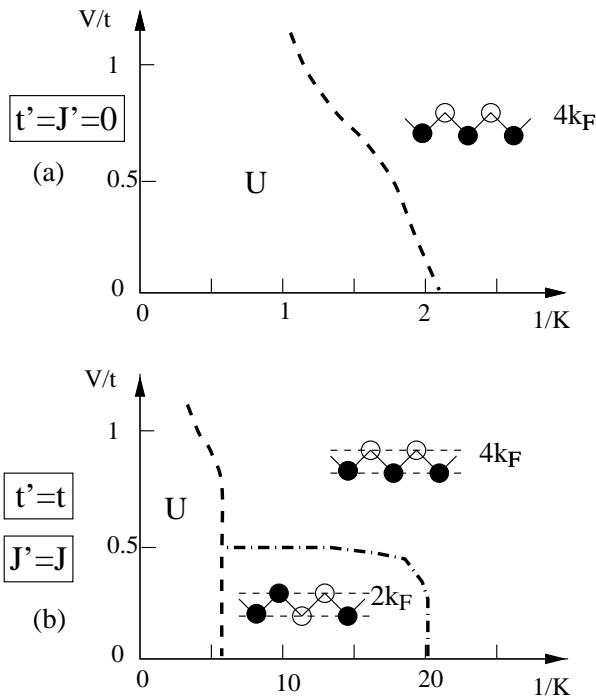,width=8truecm,angle=0}
\end{center}
\caption{
Typical phase diagrams of a $\frac{1}{4}$-filled t--J chain
for $J=0.3$ as a function of the NN repulsion $V$ and the strength
of the on-site Holstein coupling $1/K$  
obtained from ED of small periodic chains. 
A hopping $t'=t$ and a magnetic exchange $J'=J$ between NNN sites 
have been included (omitted) in (b) (in (a)). 
}
\label{tJ_Holstein}
\end{figure}

The phase diagram of Fig.~\ref{tJ_Holstein}(a) corresponding to the
$J'=t'=0$ case confirms the stabilization of the $4k_F$
charge order by the repulsion $V$. It can be seen that, similarly
to what happens in the t--J ladder case 
(Fig.~\ref{ladder_Holstein}(a)), there is a phase of charge
disproportionation occurring in the region of large Peierls
coupling $K^{-1}$, with a definite pattern.
Some important qualitative changes in the phase diagram are introduced 
by the NNN hoppings and exchange interactions $t'$ and $J'$ as
seen in Fig.~\ref{tJ_Holstein}(b) with the appearance of the 
bond-centered $2k_F$-CDW phase stable at intermediate 
Holstein coupling. As in the Hubbard chain, this phase
is suppressed by an intermediate NN Coulomb repulsion $V$. 

\section{Trellis lattice: charge instability}

\begin{figure}
\begin{center}
\psfig{figure=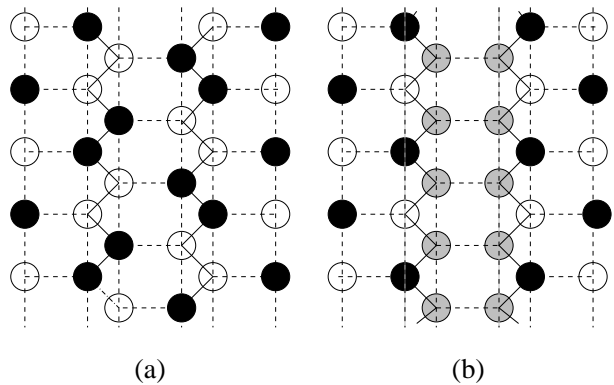,width=8truecm,angle=0}
\end{center}
\caption{
Typical patterns of charge ordering on the NaV$_2$O$_5$ lattice
structure showing a supercell of 4 sites along the zig-zag chain. The
filled (open) symbols correspond to an excess (depression) of charge 
compared to the average charge of 1/2 (grey sites have a density close
to the average density). 
(a) Bond-centered charge density wave 
of period 4 ($Q=2k_F$) with two types of non-equivalent sites.
(b) Site-centered charge density wave 
of period 4 ($Q=2k_F$) with three types of non-equivalent sites.
}
\label{2Dplane}
\end{figure}

At this stage, it is interesting to apply our results (valid,
strictly speaking, in the case of {\it isolated} chains or ladders)
to investigate the possibility of charge ordering in
the 2D trellis lattice of NaV$_2$O$_5$ (see Fig.~\ref{NaVO}(a)).
Obviously, the type of order the most likely to appear depends whether 
the chains or the ladders are the main structures of this compound.
Since the interladder couplings, although still not well determined,
are quite likely to be smaller than the couplings along the rungs of
the ladders it is to be expected that the physics of the ladders
is going to dominate the behavior of the trellis lattice. However,
this possibility has not been proven so far. Thus, here we shall
rather try to determine the most probable 2D order {\it compatible}
with both the chain and the ladder physics. A $4k_F$ charge order in
the zig-zag chains would imply the 2D formation of parallel chains of
average charge $0.5+\Delta n$ and $0.5-\Delta n$.
This corresponds to the
structure originally proposed by Galy and coworkers~\cite{Galy} 
(assuming a complete disproportionation $\Delta n=0.5$). However, 
experimental evidences accumulate in favor of a modulation of the
charge also in the x direction of the ladder legs, hence more
compatible with the $2k_F$-CDW orders in the zig-zag chains with
NNN couplings. 
Two typical patterns exhibiting bond-centered or site-centered CDW
along the zig-zag chains are shown in 
Figs.~\ref{2Dplane}(a)~and~\ref{2Dplane}(b) 
respectively. Note that only the structure of Fig.~\ref{2Dplane}(a)
is fully compatible with the zig-zag CDW ordering in {\it all} the 
ladders. Based on our previous studies, we believe this structure
should be stable in the 2D trellis lattice for large on-site 
Coulomb repulsion $U$, intermediate short-range Coulomb repulsion 
and when the hoppings and magnetic couplings along the legs
($J_\parallel= J'$, $t_\parallel= t'$) exceed the cross-bond
terms $J_{xy}$ and $t_{xy}$ (corresponding to $J$ and $t$ in the chain
model).
It is interesting to notice that, in our picture, the charge ordering
is not saturated ($\Delta n<0.5$). However, a doubling of the
periodicity occurs in the direction of the ladders for any finite value 
of the order parameter $\Delta n$ consistently with recent 
experiments~\cite{LLB_Synchrotron}.

So far, only short range Coulomb repulsion has been considered.
We briefly comment on the role of the long range Coulomb repulsion.
We have done simple classical calculations of the Madelung energy
corresponding to the various charge configuration introduced before.
As previously mentioned, most charge ordered states have lower
energies than the uniform state. Therefore, we believe that the
long range part of the Coulomb repulsion should increase further the
tendency towards charge disproportionation.

\section{Isolated chain: coexisting charge and SP orders}

We now proceed with the investigation of the last issue, namely the
role of an additional Peierls coupling in the chains.
We shall follow a method similar to the one used before for the ladder
structure. For simplicity, we use an effective potential to 
stabilize the bond-centered $2k_F$-CDW phase found previously
(see Fig.~\ref{tJ_Holstein}(b)),
\begin{equation}
H_{\rm eff} = V_{\rm eff} \sum_{i}
n_{i}\, \cos{(2\pi\frac{r_i}{4a}+\frac{\pi}{4})} \, .
\end{equation}
replacing the terms $H_V+H_H$ of Hamiltonian (\ref{tJ_chain}).
This effective potential can also include the effects of the LR
Coulomb interaction on the trellis lattice discussed in the previous
section.
The amplitude of the CDW can then be tuned by $V_{\rm eff}$.
We first consider a Peierls coupling introduced 
in the NN bonds by the following substitutions,  
\begin{eqnarray}
t\,\rightarrow\, t (1+\delta^B_{i})\, ,
\label{SP_chain1} \\
J \,\rightarrow\, J(1+2 \delta^B_{i})\, , 
\label{SP_chain2}
\end{eqnarray}
where the corresponding elastic energy has the usual form 
$\frac{1}{2}K_B\sum_{i}(\delta_{i}^B)^2$.
GS properties are obtained by solving iteratively the
set of non-linear equations,
\begin{eqnarray}
K_B\delta^B_{i} &+& 2 J \big<{\bf S}_{i}\cdot
{\bf S}_{i+1}-\frac{1}{4}n_{i}n_{i+1}\big>
\nonumber \\
&+&t \big< {\tilde c}_{i;\sigma}^\dagger{\tilde c}_{i+1;\sigma}
+ h.c.\big> =0\, .
\label{non_linear3}
\end{eqnarray}

\begin{figure}
\begin{center}
\psfig{figure=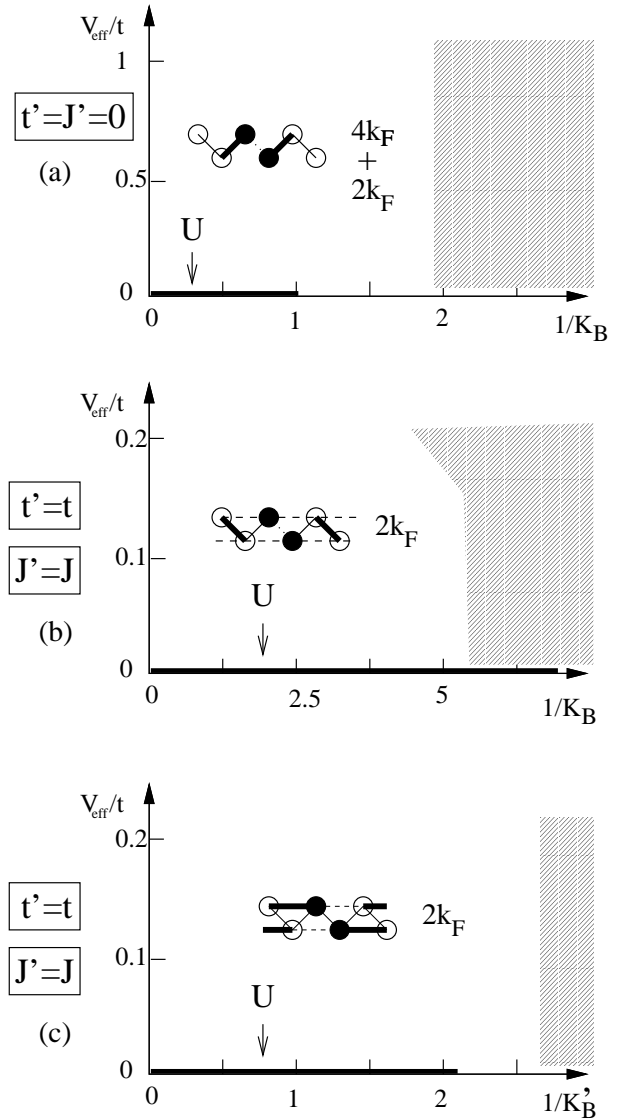,width=8truecm,angle=0}
\end{center}
\caption{
Typical phase diagrams of a $\frac{1}{4}$-filled 
t--J chain for $J=0.3$ as a function of the effective potential 
$V_{\rm eff}$ 
(see text) and the strength
of the Peierls bond couplings $1/K_B$ and $1/K_B^\prime$  
obtained from ED of small periodic chains. 
A hopping $t'$ and a magnetic exchange $J'$ between NNN sites 
have been included (resp. omitted) in (b) and (c)
(resp. in (a)). In (a) and (b) (resp. (c)) the Peierls coupling is
included only on the bonds connecting NN sites (resp. NNN sites).
}
\label{tJ_SP}
\end{figure}

\begin{figure}
\begin{center}
\psfig{figure=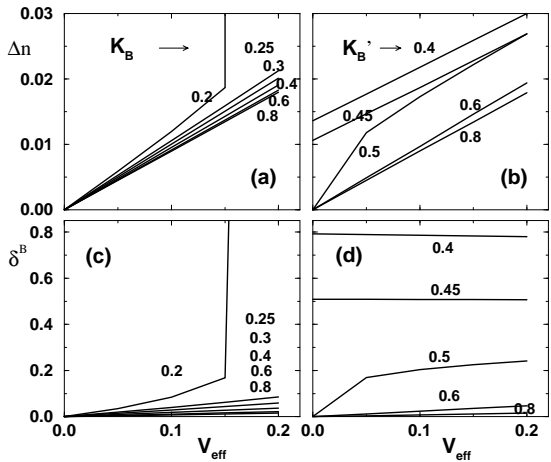,width=8truecm,angle=-90}
\end{center}
\caption{
CDW order parameter $\Delta n$ [in (a) and (b)]
and bond modulation $|\delta^B_i|$ 
[in (c) and (d)] of a $\frac{1}{4}$-filled 
t--J chain for $J=0.3$ as a function of the effective potential 
$V_{\rm eff}$ (see text) and for various lattice stiffness $K_B$ (as
indicated on the plots) obtained from ED of small periodic chains. 
A hopping $t'$ and a magnetic exchange $J'$ between NNN sites 
have been included.
In (a) and (c) (resp. (b) and (d)) the Peierls coupling is included 
on the bonds connecting NN sites (resp. NNN sites).
}
\label{tJchain_OP}
\end{figure}

Typical phase diagrams are shown in Figs.~\ref{tJ_SP}(a) and
\ref{tJ_SP}(b) for a simple t-J chain and a t-t'-J-J' chain
respectively.
Interestingly enough, a spin-Peierls instability (with the
same periodicity $\lambda_{2k_F}=4a$ as the underlying CDW) occurs for 
arbitrary small coupling $1/K_B$. This phase is characterized by
three types of bonds in the 4-sites unit cell; weak (W), strong (S) or 
intermediate (I) bonds depending on the magnitudes of $\delta^B_{i}$.
The sequence I-S-W-S in the $t'=J'=0$ case corresponds in fact to a
mixture with a $4k_F$ component. 
Note that, in the absence of $V_{\rm eff}$ and for $1/K_B>1$, a simple
dimer phase of sequence W-S-W-S (pure $4k_F$ component) is stable.
As expected, the external potential generates an additional $2k_F$
component whose magnitude increases with $V_{\rm eff}$.
It is interesting to mention here that similar orders occur e.g. in
quasi-1D organic conductors such as the SP compound (TMTTF)$_2$PF$_6$;
while the metallic phase is dimerized, a quadrupling of the 
unit cell is observed in the low temperature insulating phase.
On the other hand, in the case of the zig-zag chain with NNN couplings,
an almost pure single-Fourier ($2k_F$) CDW is observed with a S-I-W-I
sequence (Figs.~\ref{tJ_SP}(b)). It is important here to stress that,
in contrast to the ladder case studied previously, the SP order
occurs only in the presence of charge ordering. Furthermore,
the SP order parameter follows closely the magnitude of the
charge disproportionation as it is clear from
Figs.~\ref{tJchain_OP}(a) and \ref{tJchain_OP}(c). 

For completeness, we should also assume in the zig-zag chain
the possibility of an additional Peierls coupling acting
on the bonds connecting NNN sites. This effect is then characterized
by a new set of bond variables $\{\delta_i^{B\prime}\}$ and a different
elastic constant $K_B'$. In the real materials, these new parameters 
are not completely independent from the previous ones.
However, since the exact 

\noindent
relations between them, which 
depend on very fine details of the chemistry  of the material,
are quite difficult to determine,
we shall here examine the role of this new coupling
separately from the previous one. It will be easy then to add the two
effects afterwards, at least at a qualitative level. 
We then consider the following substitution,
\begin{eqnarray}
t'\,\rightarrow\, t' (1+\delta^{B\prime}_{i})\, ,
\label{SP_chain3} \\
J' \,\rightarrow\, J' (1+2\delta^{B\prime}_{i})\, , 
\label{SP_chain4}
\end{eqnarray}
for the parameters associated to the bonds connecting the NNN
sites $i$ and $i+2$. 

The set of local equations,
\begin{eqnarray}
K_{B}^\prime\delta^{B\prime}_{i} &+&2 J' \big<{\bf S}_{i}\cdot
{\bf S}_{i+2}-\frac{1}{4}n_{i}n_{i+2}\big>
\nonumber \\
&+&t' \big< {\tilde c}_{i;\sigma}^\dagger{\tilde c}_{i+2;\sigma}
+ h.c.\big> =0\, .
\label{non_linear4}
\end{eqnarray}
have been solved as before on small 12 or 16 sites periodic chains 
and the results 
are summarized in the phase diagram shown on Fig.~\ref{tJ_SP}(c).
For any arbitrary small SP coupling, a dimerization occurs in the two
parallel chains formed by the NNN bonds, coexisting with the
underlying CDW state. As seen in Figs.~\ref{tJchain_OP}(b) 
and~\ref{tJchain_OP}(d),
two qualitatively different regimes have to be distinguished;
(i) for small SP couplings (let say $1/K_B^\prime<2$), the dimerization
and the charge order appear simultaneously and follow a linear
behavior as a function of the magnitude of the external potential;
(ii) for larger couplings, the mixed CDW-SP state is stable even in the
absence of $V_{\rm eff}$. In that case, the charge disproportionation
$\Delta n$ is further increased by $V_{\rm eff}$ while the dimerization
$\delta^B$ remains almost constant. It is important to stress
here that, in both regimes, both order parameters follow a very similar
behavior as a function of the coupling constants of the model
suggesting that the two types of orders are intrinsically linked.

\section{Trellis lattice: coexisting charge and SP orders}

\begin{figure}
\begin{center}
\psfig{figure=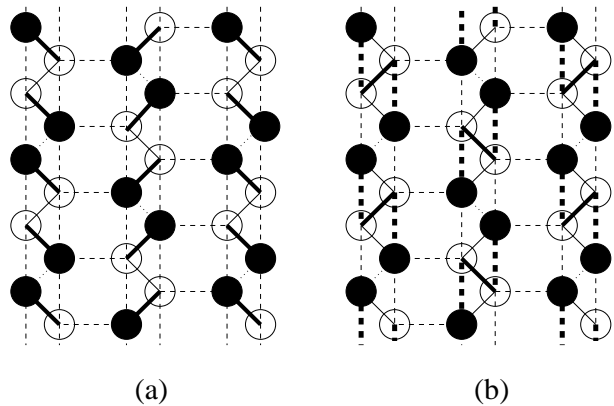,width=8truecm,angle=0}
\end{center}
\caption{
Typical patterns of coexisting lattice distortion and charge ordering 
on the NaV$_2$O$_5$ lattice structure 
showing a periodicity of 4 sites along the t-t'-J-J' chain.
The filled (open) symbols correspond to an excess (depression) of
charge compared to the average charge of 1/2. 
These patterns show a modulation of the bond exchange couplings
along the zig-zag chain 
with three types of bonds and two types of sites ($2k_F$ order of the
charge density).
Thick, thin and dotted lines correspond to strong, intermediate and 
weak bonds respectively.
(a) Mixed $2k_F$--$4k_F$ lattice distortion.
(b) $2k_F$ lattice distortion.
}
\label{2DplaneSP}
\end{figure}

Based on our previous studies of the coexistence of CDW and SP
orders in isolated ladders and zig-zag chains, we shall now
attempt to construct a unified picture for the 2D trellis lattice of
NaV$_2$O$_5$. Although this approach is not rigorous, we believe
it can provide reliable qualitative predictions of the low temperature
structure of this material.
Our starting point is the charge ordered pattern of Fig.~\ref{2Dplane},
likely to be stabilized by a local on-site Holstein coupling and/or
a LR Coulomb repulsion.
As seen before, this structure combines the zig-zag CDW pattern of the
ladders and the bond-centered $2k_F$-CDW of the chains.
If the SP instability is driven primarily by a Peierls coupling in
the diagonal bonds $t_{xy}\equiv t$ and $J_{xy}\equiv J$, one
should expect the pattern of Fig.~\ref{2DplaneSP}(a) with the
I-S-W-S sequence along the zig-zag chains. However, as argued
previously, the stability of the bond-centered $2k_F$-CDW
in the chains which we assume here 
requires rather sizable NNN couplings $t'$ and
$J'$ corresponding to the $t_\parallel$ and $J_\parallel$ 
parameters of the trellis lattice.
Therefore, we believe that the Peierls coupling
corresponding to these bonds
can no longer be neglected. In this case, one can argue that
the structure shown in Fig.~\ref{2DplaneSP}(b) is more stable
than the one shown in Fig.~\ref{2DplaneSP}(a). It is obtained by
a superposition of the two types of SP orders of the
t-t'-J-J' chains shown in Figs.~\ref{tJ_SP}(b) and \ref{tJ_SP}(c), 
each of them being mediated by Peierls couplings in the diagonal
bonds and the legs respectively. This state is characterized by a
dimerization in the legs and a modulation S-I-W-I in the t-t'-J-J'
chain.
Note that the SP order on the diagonal bonds of the t-t'-J-J' chain is
somehow ``locked" to the charge order i.e. the weak (W) bonds
are always the ones connecting two sites with an excess of charge and
so on. However, the actual pattern of the

\noindent
chains depends on the relative
phase between the two SP modulations. Fig.~\ref{2Dplane}(b)
corresponds to the simplest and most natural choice with only two
different kinds of triangles. Besides, we believe that this possibility
has the lowest energy. 
\begin{figure}
\begin{center}
\psfig{figure=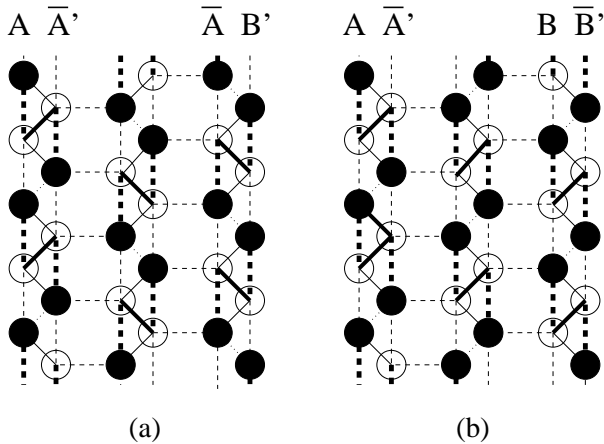,width=8truecm,angle=0}
\end{center}
\caption{
Same distortion and charge ordering pattern along
the zig-zag chains as on Fig.~\protect\ref{2DplaneSP}(b) 
but with a doubling of the periodicity in the transverse direction.
A quadrupling (doubling) of the 2D unit cell occurs in (a) ((b)).
}
\label{2DplaneSP2}
\end{figure}

The last step in our procedure consists now to refine the final
structure using our knowledge of the SP instabilities in an isolated
ladder. In fact, by examining the ladders of Fig.~\ref{2DplaneSP}(b), 
we realize that the type of dimerization 
occurring in the legs do not correspond to any type of SP order found
before in the isolated ladder. However, although still imposing a
fixed type of pattern in the legs and zig-zag chains, other
arrangements are possible in the transverse direction leading to
qualitatively different structures. Since the charge ordering doubles
the periodicity in the ladder direction,
there are two different configurations A and B of the ladder
legs, which transform into each other by a translation of a unit vector.
In addition, due to the coexisting SP-BOW order, 
a reflection with respect to a site perpendicularly to the leg leads to
two non-equivalent configurations $\bar{\rm A}$ and $\bar{\rm B}$.
As can be seen straightforwardly, the characteristic pattern of the 
zig-zag chains of Fig.~\ref{2DplaneSP}(b), can appear in 
four different ways (all related to each other by a lattice translation
and/or a reflection with respect to a site), namely 
A$\bar{\rm A'}$, $\bar{\rm A}$B', $\bar{\rm B}$A' and B$\bar{\rm B'}$
(the prime refers to the second leg). Various orders can be realized
by arbitrary sequences in the transverse direction involving different
configurations in each primitive cell.
Note that each primitive cell contains two zig-zag chains but 
the pattern in the second zig-zag chain is somehow imposed its 
two neighboring ones. The previous arrangement
shown in Fig.~\ref{2DplaneSP}(b) corresponds then to a periodic
arrangement in the transverse direction like 
-A$\bar{\rm A'}$-A$\bar{\rm A'}$- (equivalent
to e.g. -$\bar{\rm B}$B'-$\bar{\rm B}$B'- and so on). In contrast, 
the ordering patterns of
Figs.~\ref{2DplaneSP2}(a) and \ref{2DplaneSP2}(b) of the 
form -A$\bar{\rm A'}$-$\bar{\rm A}$B'- 
(equivalent to -A$\bar{\rm A'}$-$\bar{\rm B}$A'-)
and -A$\bar{\rm A'}$-B$\bar{\rm B'}$- respectively double 
the periodicity in the
transverse direction as well. It is important to stress that only the
pattern shown in Fig.~\ref{2DplaneSP2}(b) show the expected mixed
CDW-SP ordering of the D-ZZ type in {\it all} the ladders (see phase
diagrams in Figs.~\ref{ladder_Holstein}(b) and (c)). We can then
argue that the ordering shown in Fig.~\ref{2DplaneSP2}(b) can be
stabilized both by the couplings in the chains and in the ladders.
This state corresponds in fact to a check-board pattern with a
doubling of the unit cell and is consistent with recent
light and neutron scattering experiments~\cite{LLB_Synchrotron}.

\section{Conclusions}

To summarize, the role of Holstein and Peierls electron-phonon
couplings as well as magneto-phonon (SP) couplings has been
investigated in the adiabatic approximation in quarter-filled
one-dimensional chain and ladder systems. A numerical method
based on ED techniques supplemented by a self-consistent
procedure has been used to determine various phase diagrams
as a function of the strengths of the lattice couplings.
We have shown that, generically, CDW, BOW and spin-Peierls orders
coexist in such systems. Moreover, in many cases, the SP instability
is enhanced by a charge disproportionation.
Eventually, we have considered the case of the trellis lattice of
NaV$_2$O$_5$. Based on the previous results for isolated chains
and ladders, we have proposed a mixed CDW-BOW-SP ground state with
a zig-zag charge pattern in the ladders and a modulated bond structure
involving 7 different bonds. Such ordered state doubles the unit cell
introducing a check-board type pattern.

Whether the charge ordering and the BOW-SP transition occur
simultaneously at the same temperature is not yet clear but could be
resolved experimentally by thermodynamical measurements. In addition,
such transitions should have different spectroscopic signatures
due to profound changes in the GS excitation spectrum. On one hand,
charge ordering is expected to strongly enhance the charge gap
(although the insulating character of the material at high temperature
could be accounted for by a local on-site repulsion alone).
On the other hand, the opening of a spin gap should be associated to 
the SP-BOW transition.

Very recent X-ray diffuse scattering studies of
NaV$_2$O$_5$~\cite{Ravy} suggest that 3D structural fluctuations are
important above the transition and, hence, that transverse interactions
(other than
magnetic) should be important to stabilize the ordered phase. 
In our description, the SP mechanism is intrinsically linked to
the occurrence of charge ordering, possibly induced by a coupling
with the 3D lattice (treated here in the adiabatic approximation).
Therefore, we believe that the mechanism proposed in the present work
is, to some degree, weakly dependent on the actual magnetic couplings
of the trellis lattice, which, so far, are not precisely known.

\section{acknowledgements}

D.~P. and J.~R. thank IDRIS, Orsay (France) for
allocation of CPU time on the C94 and C98 Cray supercomputers. 
J.~R. acknowledges partial support from the Ministry of Education 
(France) and the Centre National de la Recherche Scientifique.

\end{document}